\documentclass{article}
\usepackage{spconf,amsmath,graphicx, color}

\usepackage[bottom=0.8in, top=0.56in, left=0.76in,right=0.76in]{geometry}

\title{Chroma Intra Prediction with attention-based CNN architectures}
\name{Marc Gorriz Blanch$^{\star}$$^{\dagger}$, Saverio Blasi$^{\star}$, Alan Smeaton$^{\dagger}$, Noel E. O’Connor$^{\dagger}$, Marta Mrak$^{\star}$\thanks{The work described in this paper has been conducted within the project JOLT funded by the European Union’s Horizon 2020 research and innovation programme under the Marie Skłodowska Curie grant agreement No 765140.}}

\address{
$^{\star}$British Broadcasting Corporation, London, UK \\
$^{\dagger}$Dublin City University, Dublin, Ireland}

\begin{document}
%
\maketitle
\vspace{-1\baselineskip}

\begin{abstract}

Neural networks can be used in video coding to improve chroma intra-prediction. In particular, usage of fully-connected networks has enabled better cross-component prediction with respect to traditional linear models. Nonetheless, state-of-the-art architectures tend to disregard the location of individual reference samples in the prediction process. This paper proposes a new neural network architecture for cross-component intra-prediction. The network uses a novel attention module to model spatial relations between reference and predicted samples. The proposed approach is integrated into the Versatile Video Coding (VVC) prediction pipeline. Experimental results demonstrate compression gains over the latest VVC anchor compared with state-of-the-art chroma intra-prediction methods based on neural networks.

\end{abstract}

\begin{figure*}[htbp]
\vspace{-1\baselineskip}
\centerline{\includegraphics[height=0.37\textwidth]{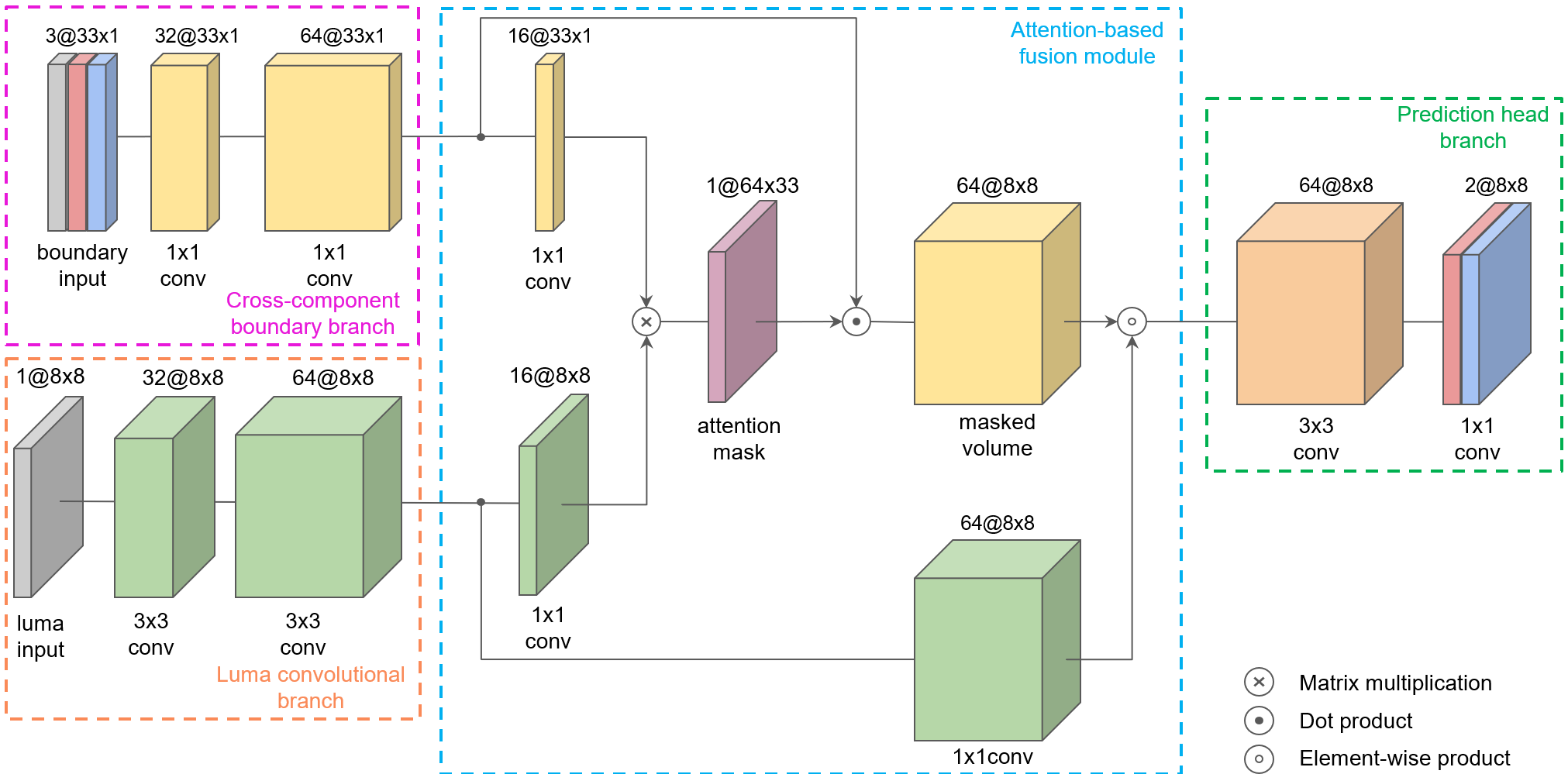}}
\vspace{-0.2\baselineskip}
\caption{Proposed architecture including the attention module used to fuse the output of the two first network branches.}
\vspace{-0.3\baselineskip}
\label{architecture}
\end{figure*}

\vspace{-0.5\baselineskip}

\section{Introduction}
\label{sec:intro}

\vspace{-0.3\baselineskip}

In the multimedia streaming era, efficient video compression has become an essential asset for tackling the increasing demand for higher quality video content and its consumption on multiple devices. New compression techniques have been developed with the aim to compact the representation of video data by identifying and removing spatial-temporal and statistical redundancies within the signal. This results in smaller bitstreams, enabling more efficient storage and transmission as well as distribution of content at higher quality with reduced resources.

Among the fundamental blocks of typical video coding schemes, intra prediction exploits spatial redundancies within a frame by predicting samples of the current block from already reconstructed samples in its close surroundings. The latest draft of the Versatile Video Coding (VVC) standard \cite{bross2019vvcdraft} (referred to as VVC in the rest of this paper) allows a large number of possible intra modes to be used on the luma component, including up to 67 directional modes and other advanced methods, at the cost of a considerable amount of signalling data. Conversely, to limit the impact of mode signalling on compression performance, a reduced number of modes is employed to intra-predict chroma samples, including the Planar, DC, pure horizontal and pure vertical modes, and the Derived Mode (DM, corresponding to using the same mode used to predict the collocated luma block). In addition, VVC introduced usage of the Cross-Component Linear Model (CCLM, or simply LM in this paper) intra modes. When using CCLM, the chroma component is predicted from the already-reconstructed luma samples using a linear model. Usage of LM prediction is effective in improving the efficiency of chroma intra-prediction. Nonetheless, the effectiveness of simple linear predictions can be limiting, and as such improved performance can be achieved using more sophisticated Machine Learning (ML) mechanisms \cite{li2018hybrid, pfaff2018intra}. Differently than these previous methods where neighbouring references are used regardless of their location, this paper proposes a new ML-based cross-component intra-prediction method which is capable of learning the spatial relations between reference and predicted samples.

A new attention module is proposed, to control the contribution of each neighbouring reference sample when computing the prediction of each chroma pixel in the current block sample location, effectively modelling the spatial information in the cross-component prediction process. As a result, the proposed scheme better captures the relationship between the luma and chroma components, resulting in more accurate prediction samples. 

\vspace{-0.6\baselineskip}

\section{Background}
\label{sec:background}

\vspace{-0.3\baselineskip}

The recent emergence of deep learning methodologies, and the impact of these new techniques in computer vision and image processing, have enabled the development of novel intelligent algorithms outperforming many state-of-the-art video compression tasks. In particular in the context of intra-prediction, a new algorithm \cite{pfaff2018intra} was introduced based on fully connected layers and Convolutional Neural Networks (CNNs) to map the prediction block positions from the reconstructed neighbouring samples, achieving BD-rate (Bjontegaard Delta rate )\cite{bjontegaard2001calculation} savings of up to 3.0\% on average over HEVC, for about 200\% increase in decoding time. The successful integration of CNN-based methods for luma intra-prediction into existing codec architectures has motivated the research of alternative methods for chroma prediction, exploiting cross-component redundancies similarly to existing LM methods. 

A novel hybrid neural network for chroma intra prediction \cite{li2018hybrid} was recently introduced. A first CNN was designed to extract features from reconstructed luma samples. This was combined with another fully-connected network used to  extract cross-component correlations between neighbouring luma and chroma samples. The resulting architecture was able to derive complex non-linear mappings for end-to-end predicting the Cb and Cr channels, but on the other hand, such approaches typically disregards the spatial location of boundary samples while predicting specific locations of the current block. To this end, an improved cross-component intra-prediction model based on neural networks is proposed, as illustrated in the rest of this paper.


\vspace{-0.4\baselineskip}

\section{Proposed method}
\label{sec:method}

\vspace{-0.3\baselineskip}

\begin{figure}[htbp]
\centerline{\includegraphics[height=0.3\textwidth]{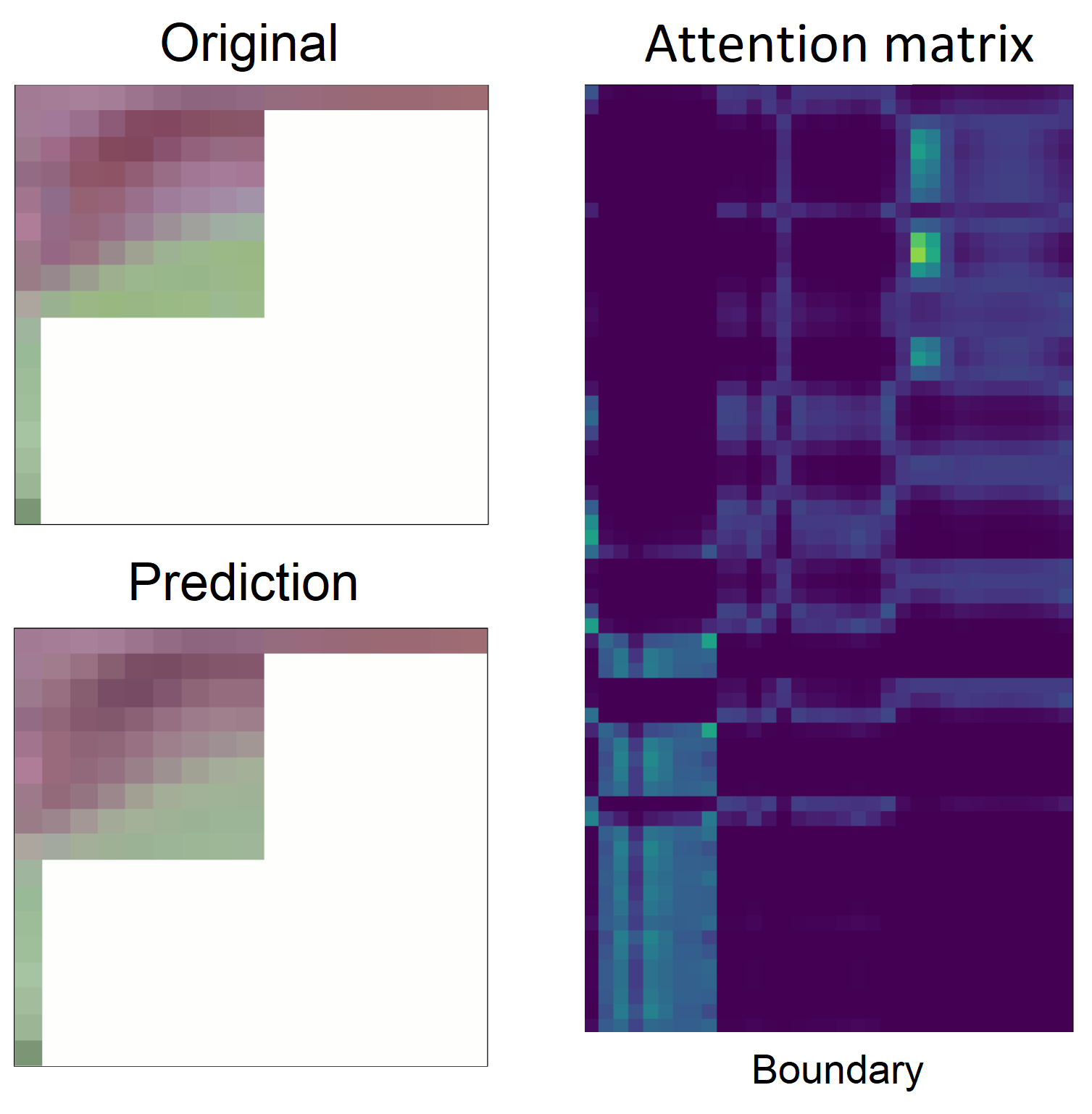}}
\vspace{-0.2\baselineskip}
\caption{Attention visualisation when predicting a block. Axes $y$ represents the $N^{2}$ block locations and axes $y$ the $B$ positions.}
\label{other-scheme}
\vspace{-0.3\baselineskip}
\end{figure}

Similarly to the model in \cite{li2018hybrid}, the proposed method adopts a scheme based on three network branches that are combined to produce prediction samples. The first two branches work concurrently to extract features from the available reconstructed samples, including the already reconstructed luma block as well as the neighbouring luma and chroma reference samples. The first branch (referred to as cross-component boundary branch) aims at extracting cross-component information from neighbouring reconstructed samples, using an extended reference array on the left of, and above the current block, as illustrated in Fig. \ref{other-scheme}. The second branch (referred to as luma convolutional branch) extracts spatial patterns over the collocated reconstructed luma block applying convolutional operations. The features from the two branches are fused together by means of an attention model, as detailed in the rest of this section. The output of the attention model is finally fed into the third network branch, to produce the resulting Cb and Cr predictions. 

An illustration of the proposed network architecture is presented in Fig. \ref{architecture}. Without loss of generality, only square blocks are considered in the rest of this section. After intra-prediction and reconstruction of a luma block, its samples can be used for prediction of collocated chroma components. In this discussion, the size of a luma block is assumed to be $N \times N$ samples, same as the size of the collocated chroma block. This may require the usage of conventional downsampling operations, for instance in the case of using chroma sub-sampled picture formats.

For the chroma prediction process, the reference samples used include the collocated luma block $X\in {\rm I\!R}^{N \times N}$, and the array of reference samples $B_c\in {\rm I\!R}^{b}$, $b = 2N + 1$ on the top-left of the current block, where $c = Y$, $Cb$ or $Cr$ to refer to the three components, respectively. $B$ is constructed from the samples on the left boundary (starting from the bottom-most sample), then the corner is added, and finally the samples on top are added (starting from the left-most sample). In case some reference samples are not available, these are padded using a predefined value. Finally,  $S\in {\rm I\!R}^{3 \times b}$ is the cross-component volume obtained by concatenating the three reference arrays $B_{Y}$, $B_{Cb}$ and $B_{Cr}$.

\vspace{-0.7\baselineskip}

\subsection{Cross-component boundary branch}
\label{ssec:ccbranch}

\vspace{-0.3\baselineskip}

The cross-component boundary branch extracts cross component features from $S\in {\rm I\!R}^{3 \times b}$ by applying $I$ consecutive $D_{i}$~-~dimensional $1 \times 1$ convolutional layers to obtain the $S_{i}\in {\rm I\!R}^{D_{i} \times b}$ output feature maps. By applying $1 \times 1$ convolutions, the boundary input dimensions are preserved, resulting in a $D$-dimensional vector of cross-component information for each boundary location. $S_{i}$ can be expressed in a neural network form as:

\vspace{-0.5\baselineskip}

\begin{equation}
S_{i}\left( S_{i-1}, W_{i} \right) = \mathcal{F}\left( W_{i} S_{i-1}^{T} + b_{i} \right),\label{eq1}
\end{equation}

\noindent where $W_{i}\in {\rm I\!R}^{D_{i} \times D_{i-1}}$ and $b_{i}$ are the $i$-layer weights and bias respectively, $D_{0} = 3$, and $\mathcal{F}$ is a Rectified Linear Unit (ReLU) non-linear activation function.

\vspace{-0.5\baselineskip}

\subsection{Luma convolutional branch}
\label{ssec:convbranch}

\vspace{-0.3\baselineskip}

In parallel with the extraction of the cross component features, the reconstructed luma block $X$ is fed to a different CNN to produce feature map volumes which represent the spatial patterns present in the luma block. The luma convolutional branch is defined by $J$ consecutive $C_{j}$-dimensional $3 \times 3$ convolutional layers with a stride of $1$, to obtain the output $X_{j}\in {\rm I\!R}^{C_{j} \times N^{2}}$ feature maps from the $N^{2}$ input samples. Similar to the previous branch, a bias and a ReLU activation are applied after each convolution operation. 

\vspace{-0.5\baselineskip}

\begin{equation}
X_{j}\left( X_{j-1}, W_{j} \right) = \mathcal{F}\left(W_{j} * X_{j-1} + b_{j} \right),\label{eq2}
\end{equation}

\noindent where $W_{j}\in {\rm I\!R}^{D_{j} \times D_{j-1}}$ and $b_{j}$ are the $j$-layer weights and bias, respectively, and $X_{0}$ is the input luma block.

\vspace{-0.5\baselineskip}

\subsection{Attention-based fusion module}
\label{ssec:attmodule}

\vspace{-0.3\baselineskip}


The concept of ”attention-based” learning is a well-known idea used in deep learning frameworks, to improve the performance of trained networks in complex prediction tasks. The idea behind attention models is to reduce complex tasks by predicting smaller ”areas of attention” that are processed sequentially in order to encourage more efficient learning. In particular, self-attention (or intra-attention) is used to assess the impact of particular input variables on the outputs, whereby the prediction is computed focusing on the most relevant elements of the same sequence \cite{cheng2016long, zhang2018self}. Extending this concept to chroma intra-prediction, this paper combines the features from the two aforementioned network branches in order to assess the impact of each input variable with respect to their spatial locations. This addresses previous limitations of similar cross-component prediction techniques, which generally discard the spatial relation of the neighbouring reference and the predicted samples. The feature maps ($S$ and $X$, from ~\ref{eq1} and ~\ref{eq2}) from the first two network branches are each convolved using a $1 \times 1$ kernel, to project them into two corresponding reduced feature spaces. Specifically, $S$ is convolved with a filter $W_{F}\in {\rm I\!R}^{h \times D}$ to obtain the $h$-dimensional feature matrix $F$. Similarly, $X$ is convolved with a filter $W_{G}\in {\rm I\!R}^{h \times C}$ to obtain the $h$-dimensional feature matrix $G$. The two matrices are multiplied together to obtain the pre-attention map $M=G^{T} F$. Finally, the attention matrix $A\in {\rm I\!R}^{N^2 \times b}$ is obtained applying a softmax operation to each element of $M$, to generate the probability of each boundary location in being able to predict a sample location in the block. Each value $\alpha_{j, i}$ in $A$ is obtained as:

\vspace{-0.4\baselineskip}

\begin{equation}
\alpha_{j, i} = \frac{\exp{(m_{i, j}/T})}{ \sum_{i=0}^{{N^2}-1} \exp{(m_{i, j}/T})},\label{eq3}
\end{equation}

\noindent where $j=0,...,N^{2}-1$ represents the sample location in the predicted block, $i=0,...,b-1$ represents a reference sample location, and $T$ is the softmax temperature parameter controlling the smoothness of the generated probabilities, with $0 \leq T \leq 1$. Notice that the smaller the value of $T$, the more localised are the obtained attention areas resulting in correspondingly less boundary samples contributing to a given prediction location, as further illustrated in Section \ref{sec:results}.

The weighted sum of the contribution of each reference sample in predicting a given output sample at specific location is obtained by computing the dot product between the cross-component boundary features $S$ (Eq. \ref{eq1}) and the attention matrix $A$ (Eq. \ref{eq3}), or formally $\langle S, A \rangle$, where $\langle . \rangle$ is the dot product. In order to further refine $\langle S, A \rangle$, this weighted sum can be multiplied by the output of the luma branch. To do so, the output of the luma branch must be transformed to change its dimensions by means of a $1 \times 1$ convolution using a matrix $W_{\bar{x}}\in {\rm I\!R}^{D \times C}$ to obtain a transformed representation, as in:

\vspace{-0.8\baselineskip}

\begin{equation}
O = \bar{X} \odot \langle S, A \rangle,\label{eq4}
\end{equation}

\noindent where $\odot$ is the element-wise product.

\vspace{-0.4\baselineskip}

\subsection{Prediction head branch}
\label{ssec:predbranch}


The output of the attention model is further fed into the third network branch, to compute the predicted chroma samples. In this branch, a final CNN is used to map the fused features from the first two branches as combined by means of the attention model into the output Cb and Cr predicted components. The prediction head branch is defined by two convolutional layers, applying $E$-dimensional $3 \times 3$ convolutional filters and then $2$-dimensional $1 \times 1$ filters for producing the output predicted values. As can be noticed, both components Cb and Cr are obtained at once following this operation. The use of the first convolutional layer is evaluated at Table \ref{results-comparison}, observing an increase in prediction accuracy when it is applied.

\vspace{-0.4\baselineskip}

\section{EXPERIMENTAL RESULTS}
\label{sec:results}


The proposed method is integrated in VVC test model VTM 7.0 \cite{chen2018algorithmvtm7}. Only $4\times4$,  $8\times8$ and $16\times16$ square blocks are supported. The resulting module was implemented as a separate mode whose usage can be signalled in the bitstream, complementing the existing VVC chroma intra prediction methods on the supported block sizes. Moreover, 4 : 2 : 0 chroma sub-sampling is assumed, where the same downsampling filters implemented in VVC are used to downsample collocated luma blocks to the size of the corresponding chroma block.

Training examples were extracted from the DIV2K dataset \cite{timofte2017ntire}, which contains high-definition high-resolution content of large diversity. This database contains $800$ training samples and $100$ samples for validation, providing $6$ lower resolution versions by downsampling by the factors of $2$, $3$ and $4$ with a bilinear and unknown filters. For each data instance, one resolution version was randomly selected and then M blocks of $N \times N$ were chosen, making balanced sets between block sizes and uniformed spatial selections within each image. All samples were converted to YCbCr colour space and further normalised to be in the range $[0, 1]$. Networks for all targeted block sizes were trained from scratch using mean squared error loss between the predicted colour components and the ground truth data, using Adam optimiser \cite{kingma2014adam} with a learning rate of $10^{-4}$.

Several constraints were considered during the implementation process. The proposed models handle variable block sizes by adapting their architecture capacity based on a trade-off between model complexity and prediction performance. As proposed in the state-of-the-art hybrid method based on CNNs \cite{li2018hybrid}, giving a fixed network structure, the depth of the convolutional layers is the most predominant factor when dealing with variable input sizes. Table \ref{params} shows the chosen hyperparameters with respect to the input block size. On the other hand, the dimension parameter $h$ within the attention module was set to $16$ for all the trained models, following a trade-off between performance and complexity. Finally, the softmax temperature $T$ was cross-validated into $T = 0.5$, ensuring a suitable balance between informative samples and noisy ones from the boundary locations. Trained models were plugged into VVC as a new chroma prediction mode, competing with traditional modes for the supported $4 \times 4$, $8 \times 8$ and $16 \times 16$ block sizes. Then, for each prediction unit, the encoder will choose between the traditional angular modes, LM models or the proposed neural network mode by minimising a rate-distortion cost criterion.

\begin{table}[h]
\label{params}
\vspace{-1\baselineskip}
\caption{Model hyperparameters per block size}
\vspace{0.4\baselineskip}
\begin{tabular}{rccc}
\hline
Branch          & $4 \times 4$    & $8 \times 8$     & $16 \times 16$ \\ \hline \hline
CC Boundary     & 16, 32          & 32, 64           & 64, 96          \\ \hline
Luma Conv       & 32, 32          & 64, 64           & 96, 96         \\ \hline
Attention       & 16, 16, 32      & 16, 16, 64       & 16, 16, 96      \\ \hline
Output          & 32, 2           & 64, 2            & 96, 2           \\ \hline
\end{tabular}
\end{table}

The proposed methodology is tested under the Common Test Conditions (CTC) \cite{ctc}, using the suggested all-intra configuration for VVC with a QP of 22, 27, 32 and 37. BD-rate is adopted to evaluate the relative compression efficiency with respect to the latest VVC anchor. Besides, a joint cross-component metric (YCbCr) \cite{bjontegaard2001calculation} is considered to evaluate the influence of the chroma gains when signalling the luma component. Test sequences include 26 video sequences of different resolutions known as Classes A, B, C, D and E. Due to the nature of the training set, only natural content sequences were considered, and screen content sequences (Class F in the CTC) were excluded from the tests. It is worth mentioning that in these tests, all block sizes were allowed to be used by the VVC encoder, including all rectangular shapes as well as larger blocks that are not supported by the proposed method. As such, the algorithm potential is highly limited, given that it is only applied to a limited range of blocks. Nonetheless, the algorithm is capable of providing consistent compression gains. The overall results are summarised in Table \ref{results-ctc}, showing average BD-rate reductions of $0.14\%$ $0.69\%$, and $0.52\%$ for Y, Cb and Cr components respectively, and an average joint YCbCr BD-rate (calculated as in \cite{combined_yuv_bd}) reduction of $0.20\%$. 

Moreover, in order to further evaluate performance of the scheme, a constrained test is also performed whereby the VVC partitioning process is limited to using only the supported square blocks of $4 \times 4$, $8 \times 8$ and $16 \times 16$ sizes. A corresponding anchor was generated for this test. Table \ref{results-sq} summarises the results for the constrained test, showing a considerable improvement over the constrained VVC anchor. Average BD-rate reductions of $0.22\%$, $1.84\%$ and $1.78\%$ are reported for the Y, Cb and Cr components respectively, as well as an average joint YCbCr reduction of $0.43\%$. In terms of complexity, even though several simplifications were considered during the integration process, the proposed solution significantly impacts the encoder and decoder time up to 120\% and 947\% on average, respectively. Future simplifications have to be adopted in order to increase computational efficiency of the scheme. Finally, the trained models were compared with the state-of-the-art hybrid architecture \cite{li2018hybrid} with the aim to evaluate the influence of the proposed attention module. Table \ref{results-comparison} summarises the results for prediction accuracy along DIV2K test set by means of averaged PSNR. 

\begin{table}[h]
\label{results-ctc}
\vspace{-1\baselineskip}
\caption{BD-rate results anchoring to VTM-7.0}
\vspace{0.4\baselineskip}
\centerline{\begin{tabular}{r|ccc|c|}
\cline{2-5}
                               & Y       & Cb       & Cr       & YCbCr     \\ \hline
\multicolumn{1}{|r|}{Class A1} & -0.18\% & -0.84\% & -0.58\% & -0.23\% \\
\multicolumn{1}{|r|}{Class A2} & -0.13\% & -0.57\% & -0.38\% & -0.19\% \\
\multicolumn{1}{|r|}{Class B}  & -0.15\% & -0.65\% & -0.67\% & -0.21\% \\
\multicolumn{1}{|r|}{Class C}  & -0.17\% & -0.63\% & -0.41\% & -0.22\% \\
\multicolumn{1}{|r|}{Class D}  & -0.17\% & -0.63\% & -0.61\% & -0.21\% \\
\multicolumn{1}{|r|}{Class E}  & -0.08\% & -0.80\% & -0.47\% & -0.16\% \\ \hline
\multicolumn{1}{|r|}{Overall}  & -0.15\% & -0.68\% & -0.53\% & -0.20\% \\ \hline
\end{tabular}}
\end{table}

\begin{table}[h]
\label{results-sq}
\vspace{-1\baselineskip}
\caption{BD-rate results for constrained test}
\vspace{0.4\baselineskip}
\centerline{\begin{tabular}{r|ccc|c|}
\cline{2-5}
                               & Y       & Cb       & Cr       & YCbCr     \\ \hline
\multicolumn{1}{|r|}{Class A1} & -0.26\% &	-2.17\% &	-1.96\% &	-0.53\% \\
\multicolumn{1}{|r|}{Class A2} & -0.22\% &	-2.37\% &	-1.64\% &	-0.50\% \\
\multicolumn{1}{|r|}{Class B}  & -0.23\% & 	-2.00\% & 	-2.17\% & 	-0.45\% \\
\multicolumn{1}{|r|}{Class C}  & -0.26\% &	-1.64\% &	-1.41\% &	-0.44\% \\
\multicolumn{1}{|r|}{Class D}  & -0.25\% &	-1.55\% &	-1.67\% &	-0.42\% \\ 	
\multicolumn{1}{|r|}{Class E}  & -0.03\% &	-1.35\% &	-1.77\% &	-0.24\% \\  \hline
\multicolumn{1}{|r|}{Overall}  & -0.22\% &	-1.84\% &	-1.78\% &   	-0.43\% \\ \hline
\end{tabular}}
\end{table}

\begin{table}[b]
\label{results-comparison}
\vspace{-1\baselineskip}
\caption{Prediction performance evaluation (PSNR)}
\vspace{0.4\baselineskip}
\centerline{\begin{tabular}{|r|l|l|l|}
\hline
Model         & \multicolumn{1}{c|}{4x4} & \multicolumn{1}{c|}{8x8} & \multicolumn{1}{c|}{16x16} \\ \hline
Hybrid CNN \cite{li2018hybrid}  & 28.61                   & 31.47                   & 33.36                     \\ \hline
Ours without head  & 29.87                   & 32.68                   & 35.77                     \\ \hline
\textbf{Ours} & \textbf{30.23}          & \textbf{33.13}          & \textbf{36.13}            \\ \hline
\end{tabular}}
\end{table}


\vspace{-0.4\baselineskip}

\section{CONCLUSIONS}
\label{sec:conclusions}
This paper proposed to improve existing approaches for chroma intra-prediction based on neural networks, introducing  a new attention module which is capable of learning spatial relations when extracting the correlational features from the neighbouring reference samples to the block prediction samples. The proposed architecture was integrated into the latest VVC anchor, signalled as a new chroma intra-prediction mode working in parallel with traditional modes towards predicting the chroma component samples. Experimental results show the effectiveness of the proposed method, achieving a remarkable compression efficiency. As future work, a complete set of network models for all VVC block sizes aim to be implemented in order to ensure a full usage of the proposed approach leading to the promising results shown in the constrained experiment. 

\bibliographystyle{IEEEbib}
\bibliography{bibliography}

\end{document}